# Conceptual Mutation Testing for Student Programming Misconceptions

Siddhartha Prasad[a], Ben Greenman[a], Tim Nelson[a], and Shriram Krishnamurthi[a]

a    Brown University, Providence, RI, USA

**Abstract**
**Context**    Students often misunderstand programming problem descriptions. This can lead them to solve the wrong problem, which creates frustration, obstructs learning, and imperils grades. Researchers have found that students can be made to better understand the problem by writing *examples* before they start programming. These examples are checked against correct and wrong implementations—analogous to mutation testing—provided by course staff. Doing so results in better student understanding of the problem as well as better test suites to accompany the program, both of which are desirable educational outcomes.
**Inquiry**    Producing mutant implementations requires care. If there are too many, or they are too obscure, students will end up spending a lot of time on an unproductive task and also become frustrated. Instead, we want a small number of mutants that each correspond to *common problem misconceptions*. This paper presents a workflow with partial automation to produce mutants of this form which, notably, are *not* those produced by mutation-testing tools.
**Approach**    We comb through student tests that *fail* a correct implementation. The student misconceptions are embedded in these failures. We then use methods to *semantically* cluster these failures. These clusters are then translated into *conceptual mutants*. These can then be run against student data to determine whether we they are better than prior methods. Some of these processes also enjoy automation.
**Knowledge**    We find that student misconceptions illustrated by failing tests can be operationalized by the above process. The resulting mutants do much better at identifying student misconceptions.
**Grounding**    Our findings are grounded in a manual analysis of student examples and a quantitative evaluation of both our clustering techniques and our process for making conceptual mutants. The clustering evaluation compares against a ground truth using standard cluster-correspondence measures, while the mutant evaluation examines how conceptual mutants perform against student data.
**Importance**    Our work contributes a workflow, with some automation, to reduce the cost and increase the effectiveness of generating conceptually interesting mutants. Such mutants can both improve learning outcomes and reduce student frustration, leading to better educational outcomes. In the process, we also identify a variation of mutation testing not commonly discussed in the software literature.

**ACM CCS 2012**
- **Applied computing → Education**;
- **Social and professional topics → Computing education**;

## The Art, Science, and Engineering of Programming

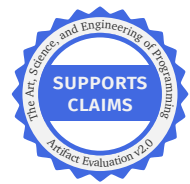

**Conceptual Mutation Testing for Student Programming Misconceptions**

## 1 Introduction

**Writing Examples** Many researchers [21, 27, 35] have observed that students often misunderstand problem statements and consequently solve the wrong problem. This not only creates frustration and leads to poor grades, but also means students may also fail to learn whatever the assignment was meant to illustrate.

In many domains, it is common to test student understanding by asking them to state the problem in their own words. Novice programmers often lack the vocabulary for this. However, multiple research teams [7, 27, 37] have observed that *test cases* can serve this purpose: they are a formal, machine-checkable restatement of the problem. Some authors [37] make a terminological distinction (which we adopt) between *tests*, which are meant to find errors in pro*gram*s, and *examples*, which are meant to explore the pro*blem*. Both use the same syntax and library support, but they are thought of differently. Tests may explore implementation edge cases, while examples—which are meant to be written before implementation even begins—probe the problem definition. Writing examples early is also emphasized in widely-used curricula such as *How to Design Programs* [10].

**Providing Feedback** If example-writing is a passive experience, however, then students neither get feedback nor have much incentive to write them. Manual feedback is hard to scale. Therefore, multiple teams have built interactive interfaces to give immediate feedback for examples [6, 7, 41]. Examplar is one such interface [37] (It does not force students to write examples first, but many do so voluntarily [38]).

Other educators have simulated such interfaces using auto-grader facilities, e.g., of systems like Gradescope (personal communication). While less elegant than in-IDE integration, it serves the same purpose of giving prompt feedback.

Examplar first runs student examples against one or more [39] correct implementations, called *wheats*. If an example fails the wheat, Examplar notifies the student that they may be on the wrong track (Figure 1a). When all examples pass the wheat, Examplar then runs a set of buggy implementations called *chaffs* to see whether the examples are thorough enough to catch the bugs. Valid examples that fail to catch the chaffs may be a symptom that the student does not fully understand the problem.

To use Wrenn's example [37], assume students are asked to implement a function to find the median. Many examples for median (e.g., median([1, 2, 3]) is 2) could just as well be examples for mean and mode. These examples pass the wheats, but do not sufficiently demonstrate that students have understood their task. Thus, it is critical that mean and mode (among others) are available as chaff implementations. If student examples do not *kill* the chaffs (borrowing terminology from mutation testing), this is a hint to hold off on implementation and instead demonstrate a more-thorough understanding of the problem by writing a broader suite of examples.

**How to Design Mutants?** Unfortunately, the Examplar project leaves open a key issue: how to design conceptual mutants (a.k.a. chaffs). If there are too many mutants, the gamified interface may tempt students into catching them all instead of solving the **(a)** Examples that fail the hidden wheat are incorrect.

**(b)** Correct examples that fail to catch all chaffs are not thorough.

**Figure 1** Examplar feedback for incorrect and incomplete examples.

assigned problem [37]. Therefore, to both be useful and not be counterproductive, mutants:
- must be few in number,
- must correspond to problem misconceptions (not random bugs), and
- should ideally correspond to misconceptions students actually have.

Whereas obscure mutants are helpful in traditional mutation testing, this educational context calls for few and fruitful mutants to avoid overwhelming students.

**Finding Misconceptions** How do we learn what misconceptions students have? The education literature has long documented the phenomenon of the "expert blind spot" [23]: experts are often poor at predicting what students will find difficult.

Prasad et al. [26] explore this phenomenon in the Examplar context. Manually-constructed chaffs, even though they had been curated by experts over several years, did not accurately predict most student misunderstandings.

If both traditional mutation testing techniques and expert predictions of student misunderstandings cannot be relied on to produce chaffs, what techniques *can* be used? Prasad et al. [26] suggests using *wheat failures* (*incorrect* examples) as a starting point. When an example fails a wheat, it could correspond to a problem misunderstanding. For example, some students (in our experience) think that the median of [1, 3, 2] is 3 (the "middle" element) and consequently write an invalid example.

Prasad et al. [26] show that wheat failures *are* a productive source of misconceptions. However, they do so through an enormous amount of manual clustering and analysis that is simply not replicable at scale. Even a small number of students (< 100) can create thousands of examples (Section 6). Several questions must, therefore, be answered to put their insight to work:

1. How to cluster wheat failures?
2. How to derive misconceptions from clusters?
3. How to realize misconceptions as chaffs?

Our contribution is to provide a workflow that addresses these questions.

## 2 Related Work

Mutation testing [1] is a venerable technique for determining the quality of a test suite. A wide range of mutation tools exist today [20], and these tools can easily generate huge numbers of mutants. These mutants, however, might be trivial, redundant [28], or equivalent to the original program [22]. Finding methods to bridge the semantic gap between generated mutants and real-world software bugs is an ongoing challenge (e.g. [12, 28]).

An important difference between classic mutation testing and our work is that we are not trying to determine the quality of an arbitrary *test suite*. Student examples differ from tests in two important ways:

1. There is no implementation yet, so the kinds of tests that probe the corner cases of a particular implementation strategy are meaningless here.
2. Example suites are not meant to be "complete" in the testing sense. We do not want students to exhaust themselves trying to pin down every possible case of the problem. Instead, we want just enough mutants to steer students in the right general direction before they start implementing a solution.

Consequently, prior work that provides feedback on student examples uses manually-constructed mutants instead of relying on mutation tools and dealing with the semantic gap [6, 7]. Such mutants are hard to come by, but have a clear semantic payoff.

A related line of prior work is identifying redundant test cases. Chetouane et al. [4] and Xia et al. [42] use *k*-means clustering to find tests that exercise similar components of a system. Our work is seeks to cluster as well, but for incorrect examples rather than test cases and based on the authors' intent (specifically, their misconceptions).

## 3 Vocabulary

We establish some terminology that we will use in the rest of the paper:

**wheat** A (definitionally) correct implementation of an assignment.

**wheat-failing example (WFE)** An example, written in the syntax of a test case, that rejects the wheat (i.e., describes incorrect behavior).

**mutant** Any incorrect implementation of an assignment.

**chaff** A mutant that corresponds to a known student misconception.

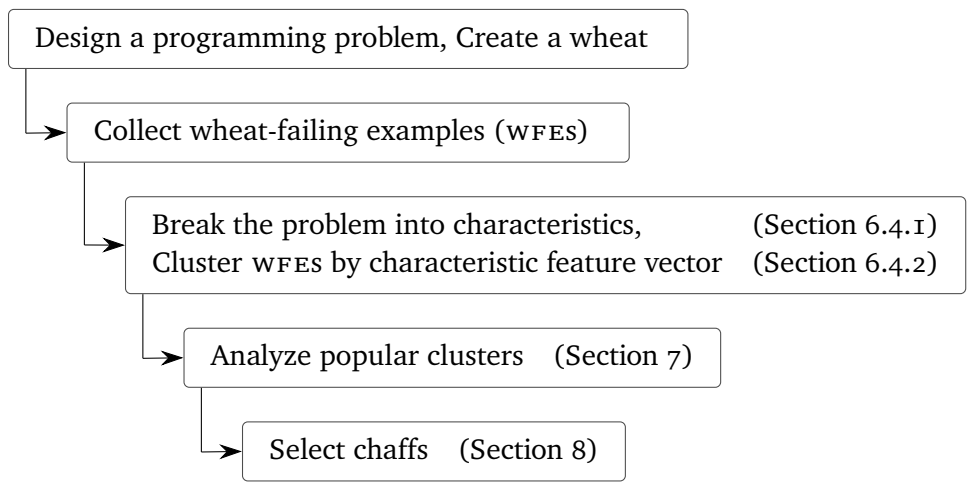

**Figure 2** How to produce a chaff suite from student data.

Traditional mutation operators are geared toward mutants rather than chaffs (Section 2); hence the workflow of this paper.

## 4 Overview

The overall workflow that we propose is shown in Figure 2. Starting from a programming problem that comes with a wheat implementation, we begin by extracting WFEs from student submissions, which is a simple matter of logging. There may be a large number of WFEs, so we need to cluster them. We employ a semantic clustering method, which requires an initial set of mutants that reflect characteristics of a correct solution (Section 6.4.1) and proceeds by running every WFE against every mutant (Section 6.4.2). We then study the clusters with the most WFEs to find common misconceptions (Section 7). The final step is to choose a few effective chaffs to deploy in the next iteration of the course (Section 8).

Each of these steps can be done manually [26], but doing so requires an infeasible amount of effort from experts (Section 6). Our additional contribution is to automate as much of the process as possible (Sections 7 and 8). We also identify research challenges that would enable further automation (Section 10).

Since the workflow begins with WFEs, we do not need any chaffs to bootstrap the process. Nevertheless, in our work, students were given an initial set of manually-constructed chaffs. This enables us to compare the chaffs developed through our workflow with those written previously, which we do in Section 9.

## 5 Study Context

We developed and tested our chaff-creation workflow in the context of three programming problems that were assigned at a highly selective, private US university as part of an accelerated introductory computer science course. The course used the Pyret programming language (https://pyret-lang.org).

1. *DocDiff* is a document similarity problem [30]. A correct solution computes the overlap between two non-empty lists of strings by reducing them to bags of words and comparing the vectors.
2. *Nile* is about an imaginary online bookstore. The problem asks for two kinds of collaborative filter: one to recommend a single book and another to recommend pairs of books.
3. *Filesystem* defines mutually-recursive datatypes to represent a Unix-like filesystem and asks for four functions: how-many, du-dir, can-find?, and fnd.

On all three assignments for three consecutive semesters, students were given access to wheats and chaffs through Examplar. The course encouraged students to write examples before starting their implementation, but did not require them to do so. The Examplar UI also encouraged the examples-first style (Figure 1a). Students were graded on both the implementation and their test suite. The latter could be checked using Examplar, creating further incentive to use it at some point during the assignment. Finally, Examplar was presented as a 24 hour teaching assistant that could answer problem definition questions immediately, automatically, and consistently, provided they could be phrased as an example [36].

**Datasets** We have data from three iterations of the course, from Fall 2020, 2021, and 2022. Each dataset plays a different role. We used the 2020 data to develop our workflow and create chaffs for 2022. This data is the focus of Sections 6 to 8. We reserved the 2021 data for evaluation, to compare our chaffs against the course staff's latest, and presumably best, chaff suites (Section 9).

During the evaluation, we discovered that the Fall 2020 data was an unfortunate training set for Nile because students used an additional tool, D4 [15], that reduced the number and character of the WFEs. Nevertheless, the chaffs that we developed using this limited dataset out-performed the 2021 chaffs (Figure 4).

## 6 Clustering WFEs

The first step in our search for common misconceptions is to cluster WFEs into a relatively small number of buckets. To this end, we introduce a semantic clustering technique in Section 6.4. We were initially hopeful, however, that existing clustering tools would suffice for WFEs. They did not; nevertheless, we report on two syntactic clustering tools in Section 6.2 and a failed attempt with the language-aware OverCode in Section 6.3.

**Why Cluster?** Clustering is critical for two reasons. First, examples are hard to analyze. By contrast to programs, most examples look fairly similar—whether or not they are correct. Each example applies a function to an input datum and makes a prediction about the result; there are few syntactic patterns to reveal their differences.

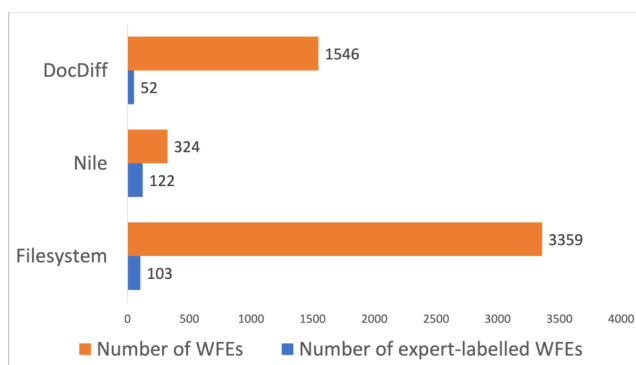

**Figure 3** Number of WFEs from 2020, number manually analysed.

Furthermore, the inputs and outputs may contain variables that point to large values spread across the codebase, adding to the cognitive load. For large programs, we can always rely on execution to show what the code does, but this approach does not work for WFEs. An example cannot run without an implementation of the function at hand. One option is to use the wheat implementation, but this gives very little insight because all WFEs fail the wheat by definition.

Second, WFEs are easy to come by. In our experience, a few dozens of students (well under 100) can produce over one thousand WFEs during one assignment (Figure 3). With three assignments and three semesters, we have too much data to analyze in reasonable time. Other educators will no doubt face the same problem.

**6.1 Ground Truth Clusters**

Using methods from grounded theory, we manually analyzed and clustered a tiny sample of the WFEs from the 2020 iteration of the three assignments (Section 5). Our prior work describes the process in detail [26].

Figure 3 counts the total number of WFEs and the number that we analyzed. There were over 4,000 WFEs in total. Of these, we analyzed 277 WFEs: 52 from DocDiff, 122 from Nile, and 103 from Filesystem. While the number analyzed may not seem large, it represents one month of work from two experts. In order to increase the ratio, we would either need more experts or more of their time. This is not a task that we can offload to non-experts (such as Mechanical Turk workers).

**Evaluation Metrics** We use two metrics to compare tool-generated clusters against the ground truth: V-Measure and Homogeneity. The V-Measure is a standard cluster analysis metric that assigns higher scores to clusters that contain only datapoints from one ground-truth class (high homogeneity) and that cover the class (high completeness) [29]. It makes no assumptions abouht underlying cluster structure. Output values range from 0 to 1. These two extremes correspond to no agreement and to perfect agreement.

Homogeneity is the component of the V-Measure that checks whether a cluster spans few ground-truth classes (ideally, one cluster should match one class). This property is particularly important for WFE clustering because each ground-truth class represents a distinct misconception. Tool-generated clusters should not span several classes; doing so would conflate misconceptions.

In sum, good clusters should have a high correspondence to the ground truth (V-Measure) and a Homogeneity score that is at least as large. (High Homogeneity should be the primary reason for the high V-Measures.) We consider a V-Measure of 0.7 or greater to be a high score, though we acknowledge that there is no universal definition of a high V-Measure.

## 6.2 Syntactic Clustering

The simplest, perhaps least-effort method for clustering is to use a syntactic approach. Such out-of-the-box methods would be of tremendous value to instructors, allowing them to easily reduce prohibitively large numbers of WFEs to manageable clusters.

There are, however, good reasons to doubt that syntactic clustering will work for WFEs (and indeed, it fares poorly). Consider the median example from earlier. The following two WFEs reflect the "middle" misunderstanding of median (Section 1), but are syntactically very different:

median([1, 3, 2]) is 3
median([4, 678, 0 , 99, 3]) is 0

Conversely, the following examples are syntactically similar, with an edit distance of 1, but represent different misunderstandings of the problem:

median([1, 2]) is 1
median([1, 2]) is 2

Both examples deal with the ambiguity surrounding even-length inputs: should the median be to the left of the middle, right of the middle, or the average of the two middle elements? The two examples above diverge. Consider two more examples:

median([5, 6, 7, 8]) is 6
median([5, 6, 7, 8]) is 7

Of the four even-length examples above, the first is semantically much closer to the third and second to the fourth, but syntactically the first and second are very close as are the third and fourth.

Nevertheless, for thoroughness, we made clusters using two standard similarity measures: Levenshtein Distance [18] and GumtreeDiff [9]. We used these measures in an affinity propagation [11] algorithm, which iteratively chose "best" WFEs to build clusters around and terminated when clusters were stable for five iterations.

For Levenshtein Distance, we obtained the following (low) scores:

◼ **Table 1** V-Measures and Homogeneity scores for Levenshtein Distance are low.

| Assignment | V-Measure | Homogeneity |
|---|---|---|
| DocDiff | 0.41 | 0.35 |
| Nile | 0.42 | 0.48 |
| Filesystem | 0.45 | 0.61 |

GumtreeDiff compares AST representations of code. It can detect insertions, deletions, moves, and even renames of source code elements. Since GumtreeDiff does not support Pyret, we implemented a very simple translation of WFEs to Python. (For the WFEs at the top of this section, only the list constructor and the is keyword needed translating.) We used GumtreeDiff's edit distances to build clusters and obtained the following (very low) scores:

■ **Table 2** V-Measures and Homogeneity scores for GumtreeDiff are low.

| Assignment | V-Measure | Homogeneity |
|---|---|---|
| DocDiff | 0.15 | 0.09 |
| Nile | 0.27 | 0.24 |
| Filesystem | 0.12 | 0.09 |

Given its AST-awareness, we were surprised that GumtreeDiff compared less favourably with ground truth than Levenshtein Distance. But since both did poorly, we did not investigate further.

### 6.3 Code Clustering: OverCode

OverCode was developed to assist educators in the visualization and analysis of large amounts of code submitted by students in MOOCs [13]. The tool utilizes both static and dynamic code analysis techniques to cluster programs.

Despite the many differences between WFEs and programs, WFEs are still code. Thus, we thought that OverCode could generate clusters. However, OverCode requires both a canonical solution and a suite of tests, neither of which are easy to define for WFEs. There is no canonical incorrect example and examples do not come with tests. We attempted to move forward with a placeholder example, a wheat implementation, and an empty test suite, but a second problem arose that prevented further progress: OverCode's dynamic analysis tracks program state. Failing examples have little state of their own to analyze—any interesting state would be inside the wheat.

### 6.4 Semantic Clustering

Given the failures of syntactic clustering and OverCode, we created a *semantic* clustering method, i.e., one that takes into account the meaning of the problem. We were loosely inspired by prior work on property-based testing [24, 40], which proposes that a property statement reflecting a problem can usefully be decomposed into a set of sub-properties whose conjunction yields the desired property.

#### 6.4.1 Extracting Problem Characteristics

Since the correctness of an example is already dictated by the wheat (and all WFEs are incorrect), our approach is to focus on the *ways that an example can be incorrect*. Therefore, we extract what we consider key characteristics in the problem statement. For example, the left column of Table 3 shows the characteristics we extracted from the problem statement of DocDiff [8], paraphrased slightly. The right column justifies

**Conceptual Mutation Testing for Student Programming Misconceptions**

▌ **Table 3** DocDiff characteristics paired with potential student mistakes.

| Characteristic | Some ways it can fail |
| --- | --- |
| Two words are the same if they have the same characters in the same order, regardless of case. | Students might consider case significant. |
| The indices of the vector correspond to words that are found in either document. The value at each index is how many times the corresponding word occurs in the document. | Students may consider the presence, rather than frequency, of each word (thereby producing a binary vector). |
| Words can be repeated in a document. | Students may choose a set instead of a list to gather words, thereby losing duplicates. |
| Normalize overlap by the squared magnitude of the larger vector. | Students may normalize by the wrong value or not normalize at all. |
| The overlap between two documents must be proportional to the dot-product of the documents' vectors. | Students may produce a binary classifier that always returns 0 or 1. |
| The overlap between pairs of documents should be treated as a float and should *not* be rounded. | Students may round to avoid problems with floating-point comparison. |
| The document vector for a document should account for *all* unique words, including those not in that document. | Students may fail to consider words from the other document, thereby producing malformed vectors (that still work with the dot-product if both documents have the same number of unique words). |

why these characteristics matter by describing corresponding ways in which students may make mistakes in their examples.

### 6.4.2 Clustering by Characteristics

Our plan is to cluster WFEs using the characteristics described above. The characteristics form the basis of a feature vector. Each WFE is given a feature vector, which is a binary vector indicating with a `m` if the WFE matches that characteristic and `d` if it does not match that characteristic. Thus, given the above seven characteristics, a WFE could have the feature vector `m m d d m d m`. We can then use standard clustering methods for binary feature vectors.

It is, however, slightly subtle to check whether a WFE matches a characteristic. The problem is that the characteristics, as stated above, are abstract statements, which need to be translated into concrete *programs*. Consider the characteristic that overlap should be normalized by the squared magnitude of the larger vector. Per the Anna Karenina principle [3], "Every passing example is essentially alike; each WFE fails in its own way". That is, there are *many ways to fail to match* the characteristic: the overlap could not be normalized at all, it could be normalized relative to the smaller vector, it could be normalized by something other than the squared magnitude, etc.

While it may be possible to use program synthesis in this setting (Section 10.5), for this work, we simply create a few different implementations that cause the characteristic to fail. Concretely, we take the wheat and manually mutate it to produce these narrowly incorrect implementations (e.g., by removing the normalization step). This results in a large number of potential chaffs — too many to present to students.

We then run each WFE against this family of potential chaffs to construct an expanded feature vector that contains an `m` for each matching chaff and a `d` for the others. Because of this expansion, feature vectors for DocDiff have 14 elements even though there are only 7 characteristics in Table 3. Similarly, Nile has 12-element feature vectors based on 8 characteristics. Filesystem has 14-element feature vectors based on 9 characteristics.

The process above creates a feature vector that indicates how each WFE corresponds to the characteristics. Now we can cluster the WFEs. We employ a straightforward clustering method: two WFEs are in the same cluster if and only if they have *exactly* the same feature vectors.

The clustering process requires multiple points of manual effort, and it is worth teasing them apart:

- One part is determining the interesting characteristics. We believe this is best done with human insight, and do not wish to automate it. Indeed, educators already perform such determinations when formalizing grading rubrics and writing down test cases for autograders. Crucially, this manual effort needs to be performed once up front; the set of characteristics can then be carried forward to future uses of the assignment. We could chose to add characteristics as we notice student mistakes (e.g., when helping them in office hours), but this is a small increment of labor.
- The other part is generating failing implementations. Again, these implementations can be reused in the future. However, especially assuming the characteristics are written in some formal language, this process is ripe for automation, as we discuss in Section 10.5.

Observe that this labor is independent of the number of students and WFEs.

**Relation to Standard Clustering Analysis** Clustering analysis aims to divide data into meaningful groups that capture the natural structure of the data [32]. In general, there are three steps to clustering: (1) identify features of the data along which to cluster, (2) use a similarity metric to compare these features, and (3) choose a clustering algorithm that uses this metric. Since similarity metrics normally yield numbers or tensors, the choice of clustering algorithm is an important and subtle point [19].

In our case, for clustering WFEs, the subtle step is the first one. An expert must decompose a programming problem into characteristics and then into mutant implementations. The later steps are simpler. Each mutant either matches or does not, leading to binary feature vectors. Our similarity metric is exact equality, and so clustering merely groups equal vectors (some may prefer to call this *classification* rather than clustering). This pipeline reflects the natural structure of our data.

### 6.4.3 Cluster Evaluation

When evaluating clustering, we first split the WFEs into two groups: those whose feature vectors are all-`d`, and those that have at least one `m`. The reason is as follows. For those that are all-`d`, our clustering method offers no insight. This could be because we did not have enough implementations (the WFE may have failed a characteristic in some unanticipated way!), or something else.

For our data, the percentage of all-`d` WFEs was 51 % for DocDiff, 71 % for Nile, and 74 % for Filesystem. We can get insight into these WFEs by checking what *manual* label they had previously been given (if any). What we found is that over 80 % of those fell into one of two categories: typo / type error, or unrelated to any misconception. This sample suggests there is not much information in these WFEs, though further sampling could uncover ideas for new candidate chaffs. It would also be useful to mechanically discover which WFEs are uninteresting, as we discuss in Section 10.3.

We now analyze the remaining 49 % (DocDiff), 29 % (Nile) and 26 % (Filesystem) of WFEs, which have at least one `m`. We have clustered these by strict equality of feature vectors. We can now compare how these clusters fare relative to the ground truth manual clustering (Section 6.1):

■ **Table 4** V-Measures and Homogeneity scores for semantic clustering are moderate-to-high.

| Assignment | V-Measure | Homogeneity |
|---|---|---|
| DocDiff | 0.83 | 0.90 |
| Nile | 0.75 | 0.82 |
| Filesystem | 0.67 | 0.85 |

These scores make two points. First, that clustering by characteristics significantly outperforms the other clustering methods we have tried. Second, it is sufficiently strong (especially in homogeneity) that we can use it as part of a workflow. This both justifies the investment in labor that this method requires, and motivates trying to better automate its steps.

## 7 Making Sense of Clusters

Semantic clustering reduces a huge set of WFEs down to a much smaller set of feature vectors. The next step is to determine *what went wrong* in each cluster that caused its examples to fail the wheats. Put another way: what part of the problem did learners misunderstand that led them to submit an incorrect example? Associating clusters with potential misunderstandings allows instructors to understand the types and frequencies of their students' misunderstandings. With this insight, they can productively adjust their assignments, lectures, and pedagogy to address these misunderstandings.

In principle, this is a labor-intensive step. Making sense of clusters is a task that calls for manual analysis by experts, ideally through several rounds of coding in which experts study a few WFEs in each cluster, pinpoint the issue in each WFE, and come up with a robust label. It is also an *interesting* step, which helps justify the labor costs. Analysis might uncover new ways of misunderstanding a problem and help improve future course offerings. However, full-fledged manual analysis over a large number of examples is not always feasible.

Thankfully, feature vectors provide a starting point for useful descriptions of the misconceptions at hand. The key is that each characteristic comes with an explanation (Section 6.4.2), thus each cluster comes with a set of explanations, one for each matching characteristic. There are four typical outcomes for these sets:

**one-m:** For a cluster with only one m, the explanation of that characteristic is a good *candidate* description of the cluster, though some WFEs may not fit that explanation.

**small-m:** For sets of 2–3 explanations, an expert is needed. There may be a misconception behind these WFEs that is subtler than the combination of 2–3 misunderstood characteristics. We recommend focusing expert time on these clusters.

**no-m:** For the cluster with no ms, our characteristics provide no help.

**large-m:** For sets of 4 or more explanations, our characteristics provide no help. But, we also advise against further analysis. The WFEs in these sets tend to have several, orthogonal issues, which makes it hard to pin down a common misunderstanding.

Sorting the clusters by size helps to further organize the space. Our approach is to sort, then focus on the small-m clusters.

**DocDiff Clusters**   For DocDiff, we identified 7 characteristics and ultimately settled on 14 chaffs for the feature vectors. Our dataset consists of 1,546 WFEs. Semantic clustering produced 64 clusters, the six largest of which are presented in Table 5. There is a long tail of smaller clusters: the median number of WFEs per cluster is 4 and there are 15 clusters that contain only one WFE.

By far the largest cluster has a feature vector with zero m's. The WFEs in this cluster do not match any of our characteristics. After a manual inspection, we concluded they are mainly due to typos.

The second-largest cluster has one matching chaff (1-m). The characteristic motivating this chaff is "The overlap between two documents must be proportional to the dot-product of the documents' vectors" (Table 3), and the actual chaff returns an overlap of zero for any inputs (which is not proportional, and obviously incorrect). Our candidate cause is therefore that students did not correctly implement the proportionality expectation. We manually analyzed half the WFEs in this cluster to determine how well they matched the candidate. Our analysis revealed that:

- some failures were due to typos rather than misunderstandings;
- some misunderstood that reordered documents can overlap;
- some misunderstood that empty strings can overlap; and
- some misunderstood other aspects of overlap.

■ **Table 5** Top six largest clusters for DocDiff 2020 and candidate descriptions.

| Size | Feature Vector | Candidate Description |
|---|---|---|
| 800 | d d d d d d d d d d d | — |
| 68 | d d d d d d d m d d d | Overlap is always 0 |
| 66 | m d d d d d d m d d d | Case sensitive |
| 52 | d d d d m d d d d d d | Normalize by $mag^4$ instead of $mag^2$ |
| 49 | d d d m d d d d d d d | Normalize by min vector |
| 48 | d d d d d d d d m d m m | — |

Going through the failing examples with this characteristic in mind helps us pin down a concrete explanation of the mistake that exemplifies this cluster: "documents have no overlap unless they are identical lists".

The third-largest cluster has two matching chaffs (2-m), which calls for an expert opinion. These chaffs are the zero-overlap chaff from the 2nd-largest cluster (discussed in the previous paragraph) and a chaff that makes case-sensitive comparisons between words. We examined a sample of WFEs from this cluster and found that they all used inputs for which case-sensitivity implied a zero overlap, e.g.:

  overlap(["A"], ["a"]) is 0

Thus, this cluster is better explained by students mistakenly thinking that case is significant, when in fact it is not.

As Section 6.4.2 indicates, there are multiple ways to incorrectly normalize word differences relative to document size. The fourth and fifth clusters have 1-m each and correspond to different ways of mis-normalizing. This shows the value of having multiple chaffs for one characteristic.

Finally, the sixth cluster has 3 matching chaffs (m's). A sample of its members revealed mostly typos.

**Nile Clusters** For Nile, we identified 8 characteristics and created 12 chaffs. The student data contains 324 WFEs. This number is much smaller than for the other two problems because, before using Examplar, students used a tool called D4 [15], which made them work through examples in the context of the data definitions. As a result, many of the issues that Examplar would have caught were pre-empted by D4.

Semantic clustering produced 12 clusters; the top 6 are in Table 6. The median cluster size is 3. Two clusters contain only one WFE.

As with DocDiff, the largest cluster has a feature vector with zero m's. We sampled WFEs from this cluster and found only typos. Sampling showed that the 1-m clusters corresponded well to the characteristic whose chaff produced that cluster. For the fourth cluster (2-m), we could not determine a misunderstanding from the examples.

**Filesystem Clusters** For Filesystem, we identified 10 characteristics and created 14 chaffs for the feature vectors. The data contains 3,359 WFEs. Semantic clustering produced 29 clusters; the top 6 are in Table 7. The median cluster size is 13. One cluster contained only one WFE.

**Table 6** Top six largest clusters for Nile 2020 and candidate descriptions.

| Size | Feature Vector | Candidate Description |
|---|---|---|
| 232 | d d d d d d d d d d | — |
| 37 | d m d d d d d d d d | Count books, not frequency |
| 23 | d d d d d m d d d d | Count pairs, not frequency |
| 9 | d d d d d d d d m m | — |
| 7 | d d d d d d d d d m | Recommend only books from 2+ collections |
| 5 | m d d d d d d d d d | Ignore case |

**Table 7** Top six largest clusters for Filesystem 2020 and candidate descriptions.

| Size | Feature Vector | Candidate Description |
|---|---|---|
| 2502 | d d d d d d d d d d d d | — |
| 232 | d d d m d d d d d d d d | can-find? always succeeds |
| 190 | d d d d d d d m d d d d | find returns duplicates |
| 63 | m m d m m m m m m m d d m | — |
| 53 | d d m d d d d d d d m d | — |
| 35 | d d d d d m d d d d d d | du-dir counts filename length |

Again, the largest cluster has zero m's and is filled with typos. There are three 1-m clusters; the chaff's characteristic was an accurate description of these. There is one 2-m cluster. As with Nile, we studied its WFEs but found no clear misunderstanding.

Most interesting of all is the cluster with a huge number of matches: 11-m out of 14 chaffs. Filesystem asks students to implement four functions (Section 5). Each function consumes the same kind of datum, which is complicated to write (instances of mutually-recursive datatypes). A brief sample of the WFEs in this cluster suggests that these relationships between functions contributed to the large number of m s. Students may have misunderstood relationships between functions (e.g., find succeeds iff can-find? does); students wrote one datum and several functions over it; etc. Consequently, we were unable to produce a description for this cluster.

## 8  Selecting Chaffs

The final step of our pipeline is to choose a small number of chaffs to expose to students via Examplar. Recall that students send examples to Examplar and receive feedback on how these perform against the chaffs (and also the wheat). The included chaffs should therefore cover common misconceptions without becoming overwhelming.

The obvious strategy is to sort clusters by size and pick *N* chaffs that match the top *N* clusters. This strategy has worked well for us with minor alterations:

- Since many WFEs are the result of typos, some of the largest clusters may not illustrate a misconception. This happened in all three of our assignments. The work-around is to skip uninformative clusters.

- A cluster can match several chaffs (it could be a 2-m or 3-m cluster). In this case, expert judgment is needed to decide whether to use some or all of the matching chaffs. One method is to manually analyze WFEs for common misconceptions. Another is to pick chaffs that match the greatest number of WFEs. (Table 9 in Section 10.1 presents per-chaff counts for our assigments.)
- When one programming problem consists of several parts, such as Filesystem, an early (or difficult) part can gather many more WFEs than others. In this case, it's important to pick at least one chaff for each subproblem.
- Outlier students can skew the ordering by submitting a huge number of similar WFEs. We observed this in 2022 when a student used a script to constantly probe Examplar for feedback. These data need to be filtered before ranking clusters. This is easy to detect using student identity.

**DocDiff Chaffs**   We selected five DocDiff chaffs from the pool of 14 characteristic chaffs to use in Examplar. Four of these came from the top clusters shown in Table 5. Note that one of these clusters has a 2-m feature vector; we used both of its chaffs (one overlapped with another cluster). The fifth reflects a common issue in the sample of WFEs that we manually analyzed (Section 6.1). It also matches more WFEs (Table 9) than any other unused chaff. The following list describes the mutation in each chaff:

1. Performs a case-sensitive comparison of words
2. Always returns 0 (no overlap)
3. Normalizes overlap based on the smaller document
4. Normalizes overlap by document magnitude rather than squared magnitude
5. Returns 1 if one document subsumes the other

**Nile Chaffs**   We selected six chaffs from the 12 candidates in the feature vectors. Four came from the top 1-m clusters. The fifth and six illustrate one issue in two contexts, because Nile has two subproblems (recommend books, recommend pairs of books) and we saw similar issues in both parts:

1. Counts number of books instead of their frequency
2. Counts number of pairs instead of their frequency
3. Recommends only books that appear in multiple collections
4. Performs case-insensitive comparisons of books
5. Recommends at most one book
6. Recommends at most one pair of books

**Filesystem Chaffs**   We chose five chaffs from the 14 candidates. Two came from the top clusters (Table 7) and two had high WFE counts (Table 9). We hand-picked the final chaff for coverage because Filesystem is a four-part problem and the other chaffs addressed only three parts of it:

1. can-find? always returns true
2. can-find? skips the root directory

3. du-dir ignores the length of file and directory lists
4. how-many counts files (good) and directories (bad)
5. find returns the current directory on failure

The careful reader will notice that there are three 1-m clusters in Table 7. And yet, we used only two. There is no deep reason for the omission: we were under pressure to deliver chaffs, could not analyze the WFEs in that cluster, and chose a different chaff that matched a greater number of WFEs (Table 9). In retrospect, we would have used all three 1-m clusters.

## 9 Evaluating Chaff Suites

We constructed chaffs using data from 2020 and deployed them in 2022. We evaluate the chaffs using a variety of data from 2020, 2021, and 2022. The same three problems were used in every year, and the problem statements stayed largely unchanged across years. The student populations between 2021 and 2022 were very similar, but due to unique circumstances from COVID-19, the 2020 population was fairly different: there were nearly twice as many students and, on average, they had somewhat less prior computing experience. Thus, it was not a given that chaffs produced from the 2020 population would work well for the 2022 students.

**Do the Chaffs Explain Errors?**  One useful way to evaluate chaff suites is to examine how they fare against new (2022) WFEs. What we would hope to see is a very small, but non-zero, number of m s in the popular feature vectors. If there are zero, then the WFEs are not captured by any chaff; if there are many, then we cannot pinpoint what is wrong with the WFEs. In either case, we do not get insight into the wheat failure. If, however, the vector has a low number of m's—say one or two—then there is a good chance that those chaffs explain why the example failed the wheat. We present the data in raw numbers and summarized in a graph (Figure 4).

The reader will note the odd outlier of Nile in 2020, which had *no* WFEs that passed a chaff. This is explained by the D4 tool [15]. We return to this issue in Section 10.3.

Statistical tests confirm that the apparent differences in the data are indeed significant. Table 8 presents p values from a two-tailed Z test comparing the mean occurrence rates of 1-m and 2-m WFEs versus other outcomes. The p values are low for the 2022 chaffs compared to prior years and high for the 2021 vs. 2020 comparison, which indicates a significant improvement. The one exception is Nile, which improved in 2021 due to D4 (Section 10.3). For completeness, Table 8 reports the actual Z scores from the two tailed test and effect sizes (Cohen's D) with confidence intervals.

We have two remarks about the data. First, recall from Section 6.4.3 that a large portion of WFEs are non-semantic mistakes like typos. Thus, we should not be surprised that about 70 % of WFEs are not characterized by chaffs, and can assume a ceiling of about 30 %. Second, one of our goals was to overcome the expert blind spot. The chaff suites for 2020 and 2021 were produced by experts based on several years of

**Conceptual Mutation Testing for Student Programming Misconceptions**

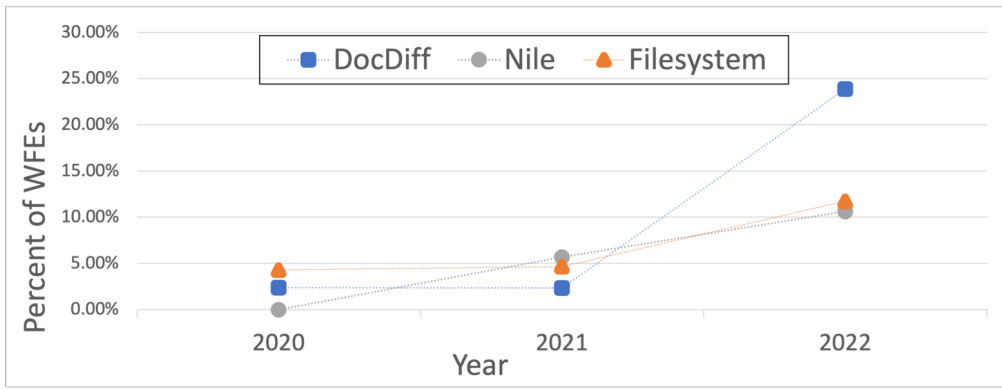

| Assignment | Year | Number of WFEs | WFEs in 1-m or 2-m clusters |
|---|---|---:|---:|
| DocDiff | 2020 | 1546 | 37 (2.39%) |
|  | 2021 | 987 | 23 (2.33%) |
|  | **2022** | **591** | **141 (23.86%)** |
| Nile | 2020 | 286 | 0 (0.00%) |
|  | 2021 | 1197 | 68 (5.68%) |
|  | **2022** | **462** | **49 (10.61%)** |
| Filesystem | 2020 | 3359 | 144 (4.28%) |
|  | 2021 | 3121 | 145 (4.65%) |
|  | **2022** | **895** | **105 (11.73%)** |

■ **Figure 4** Counting 1-m and 2-m WFEs for instructor-chosen chaffs (2020 and 2021) and chaffs produced from data (2022).

observing students working with these problems. And yet, our 2022 workflow, which incorporates student-sourced data, handily beats the experts.

**Are All Chaffs Useful?** The data above tell us that our workflow produces chaff suites that are useful in determining what is wrong with a WFE. However, there are still open questions about the nature of these chaff-suites.

1. Are some chaffs redundant? It could be that only a small number of chaffs provide value, and the remainder could have been elided.
2. Are some chaffs too general? It is possible that certain chaffs showed up in almost all feature vectors. These could be split into multiple chaffs.

After statistical analysis of the chaff suites and WFEs, we found that the data did not suggest clear answers to these questions across all the problems. The one clear improvement over previous years was that our 2022 methodology reduced chaff suite size without reducing the rates of small-m feature vectors.

■ **Table 8** Our 2022 chaffs gave 1-m/2-m outcomes significantly more often than prior chaffs. The 2021 vs. 2020 results are similar except for Nile, which used D4 in 2021.

| Matchup | Assignment | p value | Z score | Effect Size [95 % CI] (Cohen's D) |
|---|---|---|---|---|
| 2022 vs 2020 | DocDiff | 1.35E-29 | -11.24 | 0.66 [-0.75, -0.57] |
|  | Nile | 9.07E-14 | -7.36 | -0.41 [-0.55, -0.26] |
|  | Filesystem | 2.35E-10 | -6.22 | -0.28 [-0.35, -0.21] |
| 2022 vs 2021 | DocDiff | 6.87E-29 | -11.09 | -0.61 [-0.70, -0.51] |
|  | Nile | 1.82E-03 | -2.91 | -0.17 [-0.27, -0.07] |
|  | Filesystem | 2.32E-09 | -5.86 | -0.26 [-0.33, -0.19] |
| 2021 vs 2020 | DocDiff | 4.60E-01 | 0.1 | 0 [-0.07, 0.08] |
|  | Nile | 1.15E-17 | -8.48 | -0.26 [-0.39, -0.13] |
|  | Filesystem | 2.52E-01 | -0.67 | -0.02 [-0.06, 0.03] |

## 10 Discussion and Future Work

Beyond the obvious issues of trying this workflow with more problems, different students, other styles of courses, and so on, there are several salient points about the work we have presented that suggest important ways to go forward.

### 10.1 Why Rank Clusters and Not Chaffs?

Our chaff selection strategy is to rank clusters by size and then have experts draw chaffs from the top clusters (Section 8). An alternative is to skip clustering and simply rank chaffs by the number of WFEs they match (Table 9), choosing the top $N$ chaffs or perhaps the top $N$ uncovered by prior chaffs. There are two issues with this approach: it favors coarse-grained chaffs over specific ones and can reduce problem coverage.

**Redundant Chaffs** Chaffs that fail for a coarse reason typically match more WFEs than chaffs that reflect a specific issue. For example, in DocDiff, the top 5 chaffs by WFE count are the following:

1. Always returns 1
2. Rounds overlap to 0 or 1
3. Always returns 0
4. Performs a case-sensitive comparison of words
5. Returns 1 if one document subsumes the other

The top chaff is coarse; it returns 1 no matter the input. Although it matches many WFEs (280), it gives little insight as to why those WFEs are incorrect. By constrast, chaff 5 returns 1 for a specific reason. Despite being fully covered by a more-popular chaff, it is a better choice to include because it gives a precise explanation for large fraction of the general failures (47 %).

**Table 9** Number of WFEs that match each chaff in the feature vectors.

| | | | | | | | | | | | | | | |
|---|---|---|---|---|---|---|---|---|---|---|---|---|---|---|
| DocDiff | m | m | m | m | m | m | m | m | m | m | m | m | m | |
| | 148 | 117 | 0 | 126 | 57 | 92 | 71 | 85 | 131 | 186 | 280 | 195 | 120 | 111 |
| Nile | m | m | m | m | m | m | m | m | m | m | m | m | | |
| | 13 | 37 | 0 | 4 | 0 | 0 | 23 | 1 | 0 | 1 | 11 | 25 | | |
| Filesystem | m | m | m | m | m | m | m | m | m | m | m | m | m | m |
| | 177 | 165 | 127 | 374 | 136 | 137 | 178 | 160 | 308 | 139 | 157 | 87 | 127 | 163 |

**Reduced Problem Coverage**  Choosing chaffs that only match the most WFEs also falls short when dealing with assignments like Filesystem that have distinct sub-parts. The 5 largest chaffs from Table 9 for Filesystem are:

1. du-dir ignores the length of file and directory lists
2. can-find? always returns true
3. how-many counts files (good) and directories (bad)
4. how-many counts only directories
5. can-find? always returns false

These chaffs focus on the first sub-problems of the assignment and leave the final one (find) entirely untouched, thereby giving students no feedback on it.

## 10.2 From Failure to Education

This paper argues for turning student misconceptions into chaffs. Of course, misconceptions can also be used in other ways: refining the wording of assignments, changing lecture strategies, informing TAing interventions, etc. Doing so hopefully mitigates or eliminates some misconceptions, but it can also introduce new ones. As a result, the set of chaffs may have to keep changing over time to track misconceptions.

The methodology presented in this paper enables easy chaff updates year-over-year. Additionally, it allows instructors to test the effects of pedagogical interventions, seeing if and how student misconceptions change with each course iteration.

## 10.3 Challenge: Reducing Noise

As we have repeatedly found, about 70 % of student examples are typos and other low-level mistakes. These lead to large, useless clusters that a researcher nevertheless needs to wade through to confirm their lack of utility (with their lack of utility making their size especially annoying). Writing and getting no useful feedback on these is therefore likely also frustrating to students. How can we improve this situation?

One possible approach is to provide students with a structured editor to help author examples. However, such an editor would address well-formedness but not validity errors. Many low-level typos are structurally sound but just do not capture the authors intent. Given that many of the examples students construct are small, we also worry that the number of steps a required by a structured editor may prove onerous. This may dissuade students from writing examples.

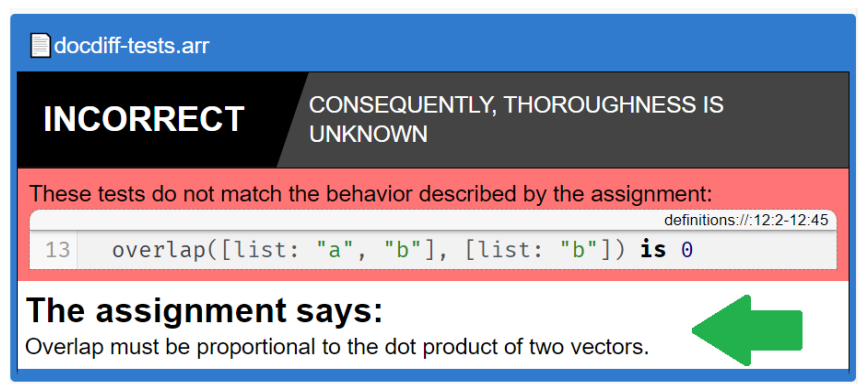

**Figure 5** Future Examplar: provide a hint (at the green arrow) for an invalid example.

The data from Nile in 2020 suggests another answer. There, we used the D4 system [15] to force students to work through *data examples* first. In the process, they effectively started Examplar with valid and thorough data. However, anecdotally, the tool proved to be rather overwhelming for some students, so we stopped its use.

Nevertheless, we believe it has the germ of a good idea: a "data Examplar" that focuses just on data instances, ignoring the *relationship* between inputs and outputs. Both "wheats" and "chaffs" would be predicates on the datatypes. Students would wrap up with a quality *data* suite, which they would then still need to arrange into the correct input-output relationships.

At that point, most of the data should be both syntactically correct and semantically valid, and the workflow here should be much more effective. Of course, we would simply have moved some of the frustration elsewhere. However, it may be *less* frustrating to think of individual data in isolation without their relationships. This would not only reduce cognitive load, but may also reduce the likelihood of frustrating, low-level phenomena like typos.

### 10.4 From Chaffs to Hints

The key benefit of getting a small, non-zero number of m s is that it opens up a new kind of student feedback. Currently, when students write a WFE, they simply get an error message about invalidity (Figure 1a). But now we can do much better: we can give a hint about their likely misconception (Figure 5).

It is up to an instructor whether they want to enable this feature or not. We believe it would be a good idea to do so, because the goal of Examplar is to get students to quickly understand the problem statement, not to play a testing game. Currently, students have to seek help from an expert to understand why their example failed. If the hints help students get on the right track quickly, they can advance quickly in their problem understanding and get to *programming* sooner and with less frustration.

The data above give us the first step towards hint-generation: the small number of m s. The next step is to confirm that the hint we would have given does in fact match the misunderstanding apparently embodied by the WFE. Our goal is to check for this and then deploy hints and measure their impact on student productivity.

Conceptual Mutation Testing for Student Programming Misconceptions

### 10.5 Challenge: Applying Program Synthesis

The task of translating problem characteristics to conceptual mutants (Section 6.4.1) has the flavor of a program synthesis problem. We start with an English sentence that describes a characteristic (e.g., "words can be repeated in a document"), observe or invent ways that students can misunderstand the characteristic (e.g., fail to track repeats), and create a mutant that reflects the misunderstanding (by modifying the wheat). This presents a challenge: to what extent can synthesis replace the (boring!) work of creating a mutant from a description of its behavior? A variant of this problem could consider the wheat and a characteristic, and mutate the wheat accordingly.

Until now, we have not tried to apply program synthesis techniques, whether using formal logic or large language models or the like. Our (limited) knowledge of the synthesis literature does not suggest the ready applicability of the techniques we know, so we present this task as a potential challenge—with working artifacts to try out—to program synthesis researchers.

A broader kind of "synthesis" question, writ large, is to generate ideas for falsifying a specification. This in turn would feed into the synthesis task described above. One possibility is to decompose the problem into subproperties and try to falsify each of these [24, 40]; this was the general inspiration for our work in Section 6.4. However, this method does not enable us to get to misconceptions like "the median is the middle element of list whether or not it's sorted." It is conceivable that more *incorrect* methods of generating content, like large language models, can actually be useful here!

### 10.6 Additional Ways to Curate Chaffs

As mentioned in Section 9, there are open questions about evaluating the value added by each chaff in a chaff-suite. We hypothesize that the ratio of wfes that passed each chaff (after accounting for outlier students) could be used as one measure of the utility of a chaff. After running all chaffs against all wfe, for the next course offering, the course staff could evaluate individual chaffs as follows:

1. If most (or all) of the wfes pass the chaff, it is under-constrained. It should probably be decomposed into multiple chaffs.
2. If most (or all) of the wfes fail the chaff, it is either over-constrained or does not reflect a common student misconception. It may be useful to remove it from the chaff suite and replace it with a more productive chaff.

### 10.7 Broader Application

What we have presented, in effect, is a workflow (that can be used iteratively) to tease out misconceptions that students have with a problem statement. This process can help to create concept inventories [17, 33]. While traditional methods for creating concept inventories are very heavyweight, reducing their burden (as we are trying to do) means they can be applied broadly, not just to large topics of broad interest [2, 5, 14, 16, 33, 34] but also to individual problems that nobody else may share.

That said, this process is agnostic to the size of the topic. The problem could be quite narrowly scoped, but it could also be about broadly-applicable ideas. DocDiff is a good example: it is about the classic TF-IDF measure [31]. Indeed, the issues we found in student data mostly reflect general misunderstandings of document similarly measure rather than any implementation details.

This suggests that the workflow we have presented here could be used in situations other than programming. After all, the very first step in Pólya's venerable *How To Solve It* [25] is to understand the problem, often by restating the problem in one's own words. Thus, one could imagine creating an Examplar-like interface for any domain that is amenable to sufficiently accurate automated checking—perhaps by creating custom user interfaces for constructing the domain examples.

## 11 Conclusion

Mutation testing for example suites is an effective way to make sure students solve the right problem, but requires a carefully curated set of mutants. Prior work has left the task of finding mutants entirely in the hands of experts, who often fail to anticipate student misconceptions [26].

Our work contributes a method to produce effective mutants at low cost by analyzing incorrect examples. The up-front cost is to decompose a correct solution into characteristics. From there, the process is partially automated: use the characteristics to build a feature vector for each incorrect example, cluster the examples with identical vectors, and derive mutants from the top-ranked vectors.

Using this process, we produced mutants in a few weeks (from thousands of incorrect examples) that out-performed expert-written mutants that had been fine-tuned over several years. The method helps experts find *semantically interesting* errors from student data which, in turn, can lead to better feedback and (eventually) better learning outcomes for students.

**Acknowledgements** Thanks to John Wrenn for helping us navigate the Examplar codebase and logs, to Nihal Nayak for introducing us to the V-Measure, and to Sreshtaa Rajesh for remembering that Nile 2020 used the D4 tool. Kuang-Chen Lu, Yanyan Ren, and Elijah Rivera provided valuable feedback on the artifact. This work was partially supported by the US National Science Foundation grant SHF-2227863 and grant 2030859 to the CRA for the CIFellows project.

Conceptual Mutation Testing for Student Programming Misconceptions

## About the authors

**Siddhartha Prasad** (siddhartha_prasad@brown.edu, https://orcid.org/0000-0001-7936-8147) is a PhD student at Brown University.

**Ben Greenman** (benjamin.l.greenman@gmail.com, https://orcid.org/0000-0001-7078-9287) is a postdoc at Brown University. He will be joining the University of Utah in Fall 2023.

**Tim Nelson** (timothy_nelson@brown.edu, https://orcid.org/0000-0002-9377-9943) preaches the good news of logic and computing at Brown University.

**Shriram Krishnamurthi** (shriram@brown.edu, https://orcid.org/0000-0001-5184-1975) is the Vice President of Programming Languages (no, not really) at Brown University.